\documentclass[12pt]{article}

\usepackage{amssymb,amsmath,bbm}
\usepackage{graphicx}
\usepackage{fullpage}
\usepackage{color}
\usepackage{hyperref}
\usepackage{longtable}

\usepackage{amssymb,amsmath}

\renewcommand{\d}{\delta}

\newcommand{\p}{\pi}
\renewcommand{\r}{\rho}
\newcommand{\s}{\sigma}\renewcommand{\S}{\Sigma}

\renewcommand{\o}{\omega}\renewcommand{\O}{\Omega}

\newcommand{\cO}{\mathcal{O}}

\newcommand{\IC}{\mathbb{C}}

\newcommand{\IF}{\mathbb{F}}

\newcommand{\IH}{\mathbb{H}}

\newcommand{\IP}{\mathbb{P}}

\newcommand{\IZ}{\mathbb{Z}}

\newcommand{\cicy}[2]{\begin{matrix} #1\end{matrix}\!\left[\begin{matrix}#2 \end{matrix}\right]}
\newcommand{\quotient}[1]{_{\hskip-2pt\lower1pt\hbox{$/$}\lower2pt\hbox{\hskip-1pt$#1$}}}
\newcommand{\ii}{\text{i}}

\def\place#1#2#3{\vbox to0pt{\kern-\parskip\kern-7pt
                             \kern-#2truein\hbox{\kern#1truein #3}
                             \vss}\nointerlineskip}

\newcommand{\sref}[1]{Section~\ref{#1}}
\newcommand{\aref}[1]{Appendix~\ref{#1}}
\newcommand{\fref}[1]{Figure~\ref{#1}}
\newcommand{\tref}[1]{Table~\ref{#1}}

\newcommand{\Dic}{\mathrm{Dic}}

\newcommand{\Tr}{\mathrm{Tr}}

\newcommand{\hodgenos}{(h^{1,1},\,h^{2,1})}
\newcommand{\symm}[1]{_{\hskip-3pt\lower3pt\hbox{$\left\{#1\right\}$}}}

\newcommand{\cicystop}{~\lower8pt\hbox{.}}

\def\place#1#2#3{\vbox to0pt{\kern-\parskip\kern-7pt
                             \kern-#2truein\hbox{\kern#1truein #3}
                             \vss}\nointerlineskip}

\newcommand{\beq}{\begin{equation}}
\newcommand{\eeq}{\end{equation}}
\newcommand{\bea}{\begin{eqnarray}}
\newcommand{\eea}{\end{eqnarray}}
\newcommand{\bean}{\begin{eqnarray*}}
\newcommand{\eean}{\end{eqnarray*}}

\newcommand{\comment}[1]{}

\newcommand{\id}{{\bf 1}}

\newcommand{\dP}{\text{dP}}

\def\str{\vrule height16pt width0pt depth10pt}
\def\Str{\vrule height21pt width0pt depth11pt}

\begin{document}
\renewcommand{\baselinestretch}{1.2}

\begin{center}
{\LARGE The Expanding Zoo of Calabi-Yau Threefolds\\}
\vspace{.5in}
Rhys Davies\footnote{\it daviesr@maths.ox.ac.uk} \\
\vspace{.15in}
{\it
Mathematical Institute, \\
University of Oxford, \\
24-29 St Giles, Oxford \\
OX1 3LB, UK}
\end{center}

\abstract{
This is a short review of recent constructions of new Calabi-Yau threefolds with
small Hodge numbers and/or non-trivial fundamental group, which are of particular
interest for model-building in the context of heterotic string theory.  The two main
tools are topological transitions and taking quotients by actions of discrete
groups.  Both of these techniques can produce new manifolds from existing ones,
and they have been used to bring many new specimens to the
previously sparse corner of the Calabi-Yau zoo where both Hodge numbers are
small.  Two new manifolds are also obtained here from hyperconifold transitions,
including the first example with fundamental group $S_3$, the smallest
non-Abelian group.  
}

\newpage

\tableofcontents

\section{Introduction}\label{sec:intro}

This paper is a short review of recent work on constructing smooth Calabi-Yau
threefolds with interesting topological properties, such as small cohomology
groups and non-trivial fundamental group.  In practice, these two properties often
go hand-in-hand, as emphasised in \cite{Triadophilia,Candelas:2008wb}.

The majority of known three-dimensional Calabi-Yau manifolds are constructed as
complete intersections in higher-dimensional toric varieties \cite{Green:1986ck,
Candelas:1987kf,Green:1987cr,Batyrev:1994pg,Kreuzer:1995cd,Kreuzer:2000xy}.
Most of the new examples found in recent years are in fact obtained from these via
one of two techniques.  The first is to take the quotient by a holomorphic action of
some finite group.  As explained in \sref{sec:quotients}, when the group action is
fixed-point-free, this is guaranteed to yield another Calabi-Yau manifold, and many
Calabi-Yau threefolds with non-trivial fundamental group have been constructed in
this way.  Several early examples can be found in \cite{Candelas:1985en,Yau1,
Strominger:1985it,Candelas:1987du,Schimmrigk:1987ke,Beauville}, but recent
efforts have brought to light many more \cite{Candelas:2008wb,GrossPopescuI,
Braun:2004xv,Batyrev:2005jc,Hua,Bouchard:2007mf,BorisovHua,GrossPopescuII,
Braun:2009qy,Braun:2010vc,Candelas:2010ve,Braun:2011hd,Bini:2011dp}, some
of which will be discussed later.  In the case that the group action has fixed points, it
is often possible to resolve the resulting orbifold singularities in such a way as to
again obtain a Calabi-Yau manifold.  Examples can be found in
\cite{Stapledon:2010mo,Freitag:2010st,Freitag:2011st}.  The second technique is to
vary either the complex structure or K\"ahler moduli of a known space until it
becomes singular, and then desingularise it by varying the other type of moduli.
Topologically, such a process is a surgery, and yields a Calabi-Yau manifold
topologically distinct from the original.  Two classes of such topological transitions,
the conifold and hyperconifold transitions, are discussed in \sref{sec:transitions},
and explicit examples of each are given.  Conifold transitions have been known
for some time to connect many Calabi-Yau threefolds
\cite{Green:1988bp,Green:1988wa,Candelas:1988di,Candelas:1989ug}, and have
been used to construct new manifolds in
\cite{Candelas:2008wb,Candelas:2010ve,Batyrev:2008rp,Filippini:2011rf}.
Hyperconifold transitions were described in \cite{Davies:2009ub}, and the first
examples of new manifolds discovered this way were given in \cite{Davies:2011is}.
The examples of \sref{sec:hyperconifolds} yield two more new manifolds, one of
which is the first known with fundamental group $S_3$, the smallest non-Abelian
group.

The fruit of these labours is that there are now many more known
Calabi-Yau threefolds with small Hodge numbers (defined arbitrarily in this paper by
$h^{1,1}+h^{2,1}~\leq~24$) than were known to the authors of \cite{Triadophilia}.  The
number with non-trivial fundamental group has also increased dramatically,
thanks largely to Braun's classification of free group actions on complete intersections
in products of projective spaces \cite{Braun:2010vc}.\footnote{The Hodge
numbers of many of these quotients are yet to be calculated, but some are likely to
be ``small" as defined above.}

The physical motivation for studying such manifolds comes predominantly from heterotic
string theory.  In this context, a non-trivial fundamental group is necessary to be able to
turn on discrete Wilson lines and thus obtain a realistic four-dimensional gauge group.
The requirement of small Hodge numbers is not so clear-cut, but it seems advantageous
if one wants to appeal to the methods of \cite{Anderson:2010mh,Anderson:2011cz} to
stabilise the moduli, and this is currently the only known way to stabilise all (geometric)
moduli in heterotic Calabi-Yau backgrounds.  Although heterotic model building is not
the theme of this review, other recent developments will be mentioned sporadically.

Throughout the paper, an arbitrary Calabi-Yau and its universal cover will be
denoted by $X$ and $\widetilde X$ respectively, while a particular Calabi-Yau
threefold with Hodge numbers $\hodgenos$ will be denoted by
$X^{h^{1,1},h^{2,1}}$.  $X^\sharp$ will denote a singular member of the family $X$,
and $\widehat X^\sharp$ a resolution of such a singular variety.

\section{Quotients by group actions}\label{sec:quotients}

\subsection{The Calabi-Yau condition}

It is an elementary fact of topology that every manifold $X$ has a simply-connected
universal covering space $\widetilde X$, from which it can be obtained as a
quotient by the free action of a group $G \cong \p_1(X)$.  We will write this relationship
as $X = \widetilde X/G$.  Although our interest is in (complex) threefolds, we will allow
the dimension $n$ of $X$ to be arbitrary through much of this section.

If $X$ is a Calabi-Yau manifold, it is easy to see that its universal cover
$\widetilde X$ is too, by pulling back the complex structure, K\"ahler form $\o$, and
holomorphic $(n,0)$-form $\O$ under the covering map (for this reason we will often
abuse notation by using the same symbols for these objects on $X$ and $\widetilde X$).
Only a little more difficult is the converse: under what conditions is $X = \widetilde X/G$
a Calabi-Yau manifold, given that $\widetilde X$ is?  There are several points to
consider (we assume always that $G$ is a finite group):
\begin{itemize}
    \item
    $X$ will be a manifold as long as the action of $G$ is fixed-point free.  Otherwise it
    will have orbifold singularities.
    \item
    It will furthermore be a complex manifold if and only if $G$ acts by biholomorphic maps.
    In this case, $X$ simply inherits the complex structure of $\widetilde X$.
    \item
    To see that $X$ is K\"ahler, pick any K\"ahler form $\o$ on $\widetilde X$.  Now note
    that for any element $g \in G$, $g^*\o$ is also a K\"ahler form, since
    $d(g^*\o) = g^*(d\o) = 0$, and for any $k$-dimensional complex submanifold $M_k$,
    \begin{equation*}
        \int_{M_k} g^*\o^k ~=~ \int_{g(M_k)} \o^k ~>~ 0~.
    \end{equation*}
    We can use this to construct a K\"ahler form which is invariant under $G$, and
    therefore descends to a K\"ahler form on $X$:
    \begin{equation*}
        \o^G ~:=~ \frac{1}{|G|}\sum_{g\in G} g^*\o~.
    \end{equation*}
    \item
    Finally, we must check whether $X$ supports a nowhere-vanishing holomorphic
    $(n,0)$-form.  This can only descend from the one on $\widetilde X$, so we need to
    check whether $\O$ is $G$-invariant.  Note that $\O$ is the unique (up to scale) element
    of $H^{n,0}(\widetilde X)$, so since $G$ acts freely, the Atiyah-Bott fixed point formula
    (\cite{AtiyahBott,AtiyahBott2}) for any $g \in G\setminus e$ reduces to
    \begin{equation*}
        0 ~=~ \sum_{q=0}^n (-1)^q\, \Tr \big( H^{n,q}(g) \big) ~=~
            \Tr\big( H^{n,0}(g) \big) + (-1)^n\, \Tr\big( H^{n,n}(g) \big)~,
    \end{equation*}
    where $H^*(g)$ denotes the induced action of $g$ on the cohomology $H^*$.
    The group $H^{n,n}(\widetilde X)$ is generated by $\big(\o^G \big)^n$, which is
    invariant, so we conclude that $\Tr\big(H^{n,0}(g)\big) = (-1)^{n+1}$, and therefore
    \begin{equation*}
        g^*\O = (-1)^{n+1}\O ~~\forall~~ g\in G\setminus e~.
    \end{equation*}
    In odd dimensions, therefore, $\O$ is automatically invariant under free group actions.
    In even dimensions, on the other hand, this simple calculation shows that there are no
    multiply-connected Calabi-Yau manifolds.
\end{itemize}
In summary, if $\widetilde X$ is a smooth Calabi-Yau threefold, then $X = \widetilde X/G$ is
a Calabi-Yau manifold if and only if $G$ acts freely and holomorphically.

\subsubsection{Smoothness}

Above, we have simply assumed that the covering space $\widetilde X$ is smooth.  In
practice, $\widetilde X$ usually belongs to a family of spaces of which only a sub-family
admits a free $G$ action.  It may be the case that, although a generic member of
$\widetilde X$ is smooth, members of the symmetric sub-family are all singular, so we
never get a smooth quotient.  Although this seems to be rare, it does occur for
$\IZ_8{\times}\IZ_8$-symmetric complete intersections of four quadrics in $\IP^7$
\cite{Strominger:1985it,Hua}, $\IZ_5{\times}\IZ_5$-symmetric complete intersections of five
bilinears in $\IP^4{\times}\IP^4$ \cite{Candelas:2008wb}, and a number
of other examples, including some found in \cite{Braun:2010vc}.  The extra condition, that
a generic \emph{symmetric} member is smooth, must therefore be checked on a
case-by-case basis.

Let us start with the special case of complete intersection Calabi-Yau manifolds in smooth
ambient spaces.  This includes what have traditionally been called the `CICY' manifolds,
where the ambient space is a product of projective spaces
\cite{Green:1986ck,Candelas:1987kf,Green:1987cr}, and certain of the toric hypersurfaces
\cite{Kreuzer:1995cd,Kreuzer:2000xy}.  The complete intersection condition means that if
the ambient space has dimension $n+k$, then the Calabi-Yau $\widetilde X$ is given by the
intersection of $k$ hypersurfaces, each given by a single polynomial equation $f_a = 0$.
In other words, the number of equations needed to specify $\widetilde X$ is equal to its
codimension.
When the ambient space is smooth, it can be covered in affine patches each isomorphic to
$\IC^{n+k}$, and the condition for $\widetilde X$ to be smooth is that
$df_1\wedge\ldots\wedge df_k$ is non-zero at every point on $\widetilde X$.  This is a very
intuitive condition --- if it holds, then at any point of $\widetilde X$ we can choose local
coordinates $x_1, \ldots,x_{n+k}$ such that $f_a = x_a + \cO(x^2)$.  Locally, then,
$\widetilde X$ projects biholomorphically onto the linear subspace
$x_1 = x_2 = \ldots = x_k = 0$, and is therefore smooth.  On any affine coordinate patch, the
components of the differential form $df_1\wedge\ldots\wedge df_k$ are just the $k{\times}k$
minors of the Jacobian matrix $J = \left(\partial f_a/\partial x_i\right)$, so the condition is that
this matrix has rank $k$ everywhere on $\widetilde X$.  It is therefore necessary to check
that there is no simultaneous solution to the equations $f_a = 0$ along with the vanishing
of all $k{\times}k$ minors of $J$, which is equivalent to the algebraic statement that the
ideal generated by the polynomials and the minors is the entire ring
$\IC[x_1,\ldots,x_{k+n}]$.  This is checked by calculating a Gr\"obner basis for the ideal,
algorithms for which are implemented in a variety of computer algebra packages
\cite{Singular,Mathematica,Gray:2008zs}; a Gr\"obner basis for the entire ring is just a
constant (usually given as $1$ or $-1$ by software).

The more general case of singular ambient spaces or non-complete intersections is not
much harder than the above.  Suppose $\widetilde X$ is not a complete intersection, so
that it is given by $l$ equations in an $n+k$-dimensional ambient space, where now we
allow $l > k$.\footnote{A typical example is the Veronese embedding of $\IP^2$ in $\IP^5$.
If we take homogeneous coordinates $z_i$ for $\IP^2$, and $w_{ij}$ for $\IP^5$, where
$j \geq i$, then the embedding is given by $w_{ij} = z_i z_j$.  The equations needed to
specify the image of this map are $w_{ij}w_{kl} - w_{il}w_{kj} = 0$, which amount to six
independent equations, whereas the codimension of the embedded surface is only
three.}  Then the condition for $\widetilde X$ to be smooth is still that the rank of the
Jacobian be equal to $k$ (the codimension) everywhere on $\widetilde X$
\cite{Hartshorne}.  The reasoning is the same as before --- if this is true, $k$ of the
polynomials will provide good local coordinates on the ambient space, allowing us to
define a smooth coordinate patch on $\widetilde X$.

If some affine patch on the ambient space is singular, it
can still be embedded in $\IC^N$ for some $N$, by polynomial equations
$F_1 = \ldots = F_K = 0$.  The Calabi-Yau is then given by
$F_1 = \ldots = F_K = f_1 = \ldots = f_l = 0$, and the condition for smoothness is once
again that the Jacobian has rank equal to the codimension, $N - n$, at all points.

For examples of interest, Gr\"obner basis calculations are often very computationally
intensive, since at intermediate stages the number of polynomials, as well as their
coefficients, can become extremely large.  It is therefore convenient to choose integer
coefficients for all polynomials, and perform the calculation over a finite field $\IF_p$.  As
explained in \cite{Candelas:2008wb}, if a collection of polynomial equations are
inconsistent over $\IF_p$, then they are also inconsistent over $\IC$, so the corresponding
variety is smooth.

Note that there does exist a slight variation on the above procedure which still leads to
smooth quotient manifolds.  It may be the case that although the symmetric manifolds
admit a free group action, they are all singular.  If, however, these singularities can be
resolved in a group-invariant way, the resolved space still admits a free group action,
with a smooth quotient.  Examples can be found in
\cite{GrossPopescuI,Hua,GrossPopescuII}.

The final possibility is that the symmetric manifolds are smooth, but the group action
always has fixed points, in which case the quotient space has orbifold singularities.
It is frequently possible to resolve these in such a way as to again obtain a
Calabi-Yau manifold, but this will not be discussed in detail here.

\subsection{Notable Examples}

\subsubsection{New Three-Generation Manifolds} \label{sec:(8,44)}

Calabi-Yau threefolds with Euler number $\chi = \pm 6$ were of particular
interest in the early days of string phenomenology, since these give physical models with
three generations of fermions via the `standard embedding' compactification of the heterotic
string \cite{Candelas:1985en}.  This typically gives an $E_6$ grand unified theory, and
although the gauge symmetry can be partially broken by Wilson lines, it is impossible to obtain
exactly the standard model gauge group in this way \cite{McInnes:1989rg}.  Nevertheless, it
was argued by Witten that deformations of the standard embedding, combined with Wilson
lines, can give realistic models \cite{Witten:1985bz}, and this was put on firmer mathematical
foundations by Li and Yau \cite{Li:2004hx}.

The archetypal example of a three-generation manifold is Yau's manifold, with fundamental
group $\IZ_3$ \cite{Yau1}, but recently two new promising three-generation manifolds were
constructed in \cite{Braun:2009qy}.  These are quotients of a manifold $X^{8,44}$ by
groups of order twelve, which are the cyclic group $\IZ_{12}$ and the non-Abelian group
$\Dic_3 \cong \IZ_3{\rtimes}\IZ_4$ (this is generated by two elements, one of order three
and one of order four, satisfying $g_4 g_3 g_4^{-1} = g_3^2$), and each has Hodge
numbers $\hodgenos = (1,4)$.  Unfortunately, it was shown in \cite{Davies:DPhil} that
the physical model on the non-Abelian quotient does not admit a deformation which yields
exactly the field content of the minimal supersymmetric standard model (MSSM) in four
dimensions.  However, the $\IZ_{12}$ quotient allows many more distinct deformations, and
the analysis of the corresponding physical models has not been completed.

\begin{figure}
\begin{center}
    \includegraphics[width=.35\textwidth]{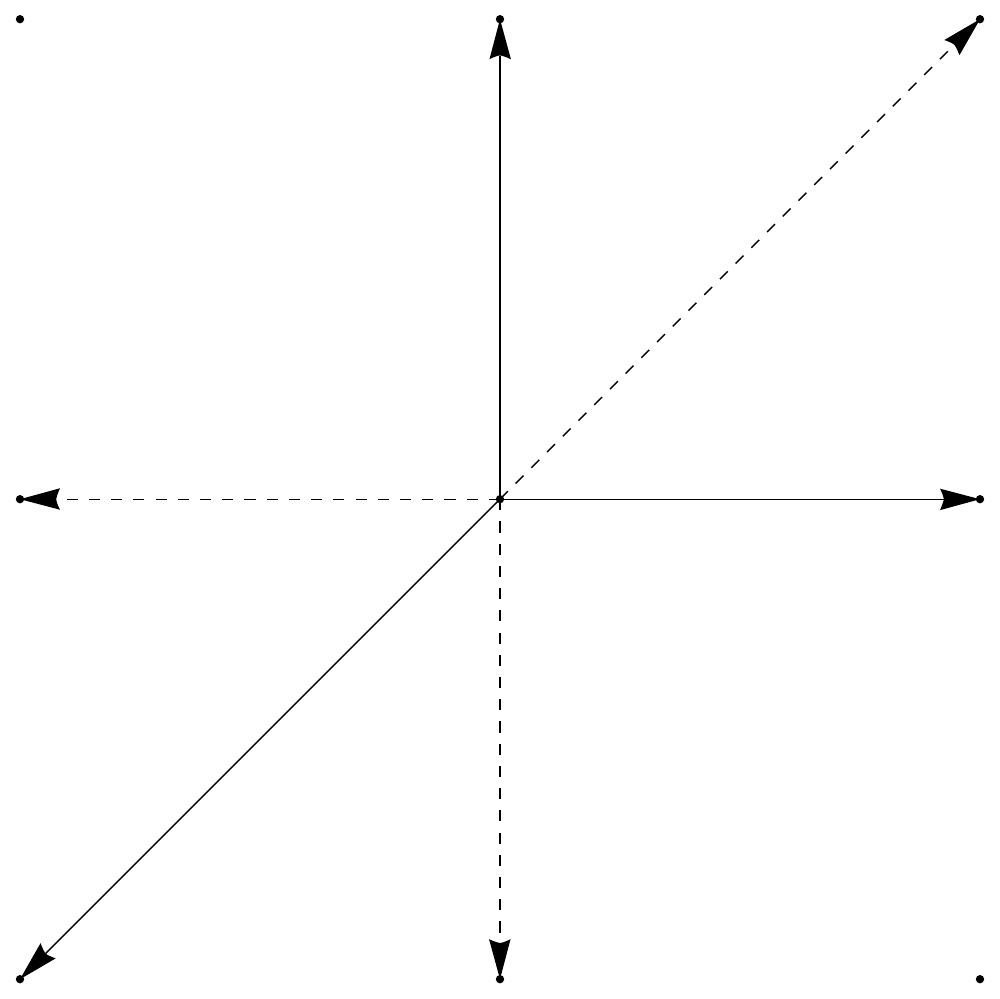}
    \parbox{.8\textwidth}{\caption{\label{fig:dP6fan}
    \small
    The fan for the toric surface $\dP_6$.  Removing the dashed rays corresponds to the
    projection to $\IP^2$.  All graphics were produced in Mathematica~\cite{Mathematica}.
    }}
\end{center}
\vspace{-15pt}
\end{figure}

The covering space $X^{8,44}$ is an anti-canonical hypersurface in $\dP_6{\times}\dP_6$,
where $\dP_6$ is the del Pezzo surface of degree six, which is $\IP^2$ blown up in
three generic points.  This surface is rigid and toric, and its fan is shown in \fref{fig:dP6fan}.

As well as the action of the torus $\big(\IC^*\big)^2$, $\dP_6$ also admits an action
by the dihedral group $D_6$, as suggested by its fan.  This can be realised as a
group of lattice morphisms preserving the fan, generated by an order-six rotation
$\r$ and a reflection $\s$, with matrix representations
\begin{equation*}
    \r = \left(\begin{array}{rr} 1 & -1 \\ 1 & 0 \end{array} \right)
    ~,~~
    \s = \left(\begin{array}{rr} 0 & 1 \\ 1 & \phantom{-}0 \end{array} \right) ~.
\end{equation*}
The product $\dP_6{\times}\dP_6$ therefore has symmetry group
$(D_6{\times}D_6)\rtimes\IZ_2$, where the extra $\IZ_2$ factor swaps the two copies
of the surface.  The quotient groups $\Dic_3$ and $\IZ_{12}$ are both order-twelve
subgroups of this which act transitively on the vertices of the fan.  Many more details
can be found in \cite{Braun:2009qy}.

\subsubsection{Quotients of the (19,19) Manifold} \label{sec:(19,19)}

The Euler number of a three-dimensional Calabi-Yau manifold is given by the
simple formula $\chi = 2(h^{1,1} - h^{2,1})$.  If a group $G$ acts freely, then
$\chi(\widetilde X/G) = \chi(\widetilde X)/G$, so this gives a simple necessary condition
for the existence of such an action: the order of the group must divide $\chi/2$.  This
usually gives a fairly strong restriction on the groups which can act freely on any given
manifold.  The only time it gives no restriction is when $\chi = 0$.
In this section we will look at a particular manifold, $X^{19,19}$, which admits free
actions by a number of disparate groups, including groups of order five, eight, and
nine.  For a Calabi-Yau threefold with $\chi \neq 0$, this would imply $|\chi| \geq 720$.

The manifold $X^{19,19}$ can be represented in a number of different ways.  Abstractly,
it is the fibre product of two rational surfaces, each elliptically fibred over $\IP^1$
\cite{Braun:2004xv,Bouchard:2007mf}.  Such
a surface is given by blowing up $\IP^2$ at the nine points given by the intersection of
two cubic curves; if we take homogeneous coordinates $t_0, t_1$ on $\IP^1$ and
$z_0, z_1, z_2$ on $\IP^2$, the corresponding equation is
\begin{equation}\label{eq:ellipticsurface}
    f(z) t_0 + g(z) t_1 = 0 ~,
\end{equation}
where $f$ and $g$ are homogeneous cubic polynomials.  We can easily see that this
corresponds to $\IP^2$ blown up at the nine points given by $f = g = 0$.  Indeed, for any
point of $\IP^2$ where $f\neq 0$ or $g\neq 0$, we get a unique solution for
$[t_0 : t_1]$, whereas for $f = g = 0$, the equation is satisfied identically, giving
a whole copy of $\IP^1$.  To see that it is also an elliptic fibration over $\IP^1$, note
that for any fixed value of $[t_0 : t_1] \in \IP^1$, we get a cubic equation in $\IP^2$, which
defines an elliptic curve.

To get the fibre product of two such surfaces, we introduce another $\IP^2$, with
homogeneous coordinates $w_0, w_1, w_2$, and another equation of the form
\eqref{eq:ellipticsurface} over the same $\IP^1$.  The resulting threefold is Calabi-Yau,
and has a projection to $\IP^1$, with typical fibre which is a product of two elliptic curves.
In \cite{Bouchard:2007mf}, Bouchard and Donagi studied group actions which
preserve the elliptic fibration, and found free actions by the groups $\IZ_3{\times}\IZ_3$,
$\IZ_4{\times}\IZ_2$, $\IZ_6 \cong \IZ_3{\times}\IZ_2$ and $\IZ_5$ (as well as all
subgroups of these, of course).

Certain of these quotient manifolds have in fact played crucial roles in the heterotic string
literature.  A model with the spectrum of the $U(1)_{B-L}$-extended supersymmetric
standard model was constructed on a quotient by $\IZ_3{\times}\IZ_3$ and studied in
\cite{Braun:2005nv,Braun:2006me} (a similar model on the $\IZ_3{\times}\IZ_3$ quotient
of the `bicubic', which is related to this manifold by a conifold transition, was found in
\cite{Anderson:2009mh}), while a model with the exact MSSM spectrum exists on a
$\IZ_2$ quotient, and was described in \cite{Bouchard:2005ag,Bouchard:2006dn}.
In \cite{Braun:2007xh,Braun:2007vy}, the quotient by a different $\IZ_3{\times}\IZ_3$
action was used as a test case for calculating instanton corrections on manifolds with
torsion curves.

There are in fact further (relatively large) groups which act freely on $X^{19,19}$, which
can be easily described using its representation(s) as a CICY.  First, we note that the
fibre product construction above is equivalent to the rather more prosaic statement that
the manifold is a complete intersection of two hypersurfaces in
$\IP^1{\times}\IP^2{\times}\IP^2$, of multi-degrees $(1,3,0)$ and $(1,0,3)$.  In the notation
of $\cite{Green:1986ck,Candelas:1987kf}$, $X^{19,19}$ can therefore be specified by the
`configuration matrix'
\vskip-10pt
\begin{equation*}
    \cicy{\IP^1 \\ \IP^2 \\ \IP^2}{
        1 & 1 \\
        3 & 0 \\
        0 & 3}\quad.
\end{equation*}
By utilising various \emph{splittings} and \emph{contractions} (see e.g.
\cite{Candelas:2008wb,Candelas:1987kf,Green:1988bp,Green:1988wa}), and checking
that the Euler number remains constant, it is easy to show that $X^{19,19}$ can also be
specified by the configuration matrices
\begin{equation*}
    \cicy{\IP^1\\ \IP^1 \\ \IP^1 \\ \IP^1 \\ \IP^1}{
        1 & 1 \\
        2 & 0 \\
        2 & 0 \\
        0 & 2 \\
        0 & 2}  \qquad, \qquad
    \cicy{\IP^1\\ \IP^2\\ \IP^2\\ \IP^2\\ \IP^2}{  
    1 & 1 & 0 & 0 & 0 & 0\\
    1 & 0 & 1 & 1 & 0 & 0 \\
    1 & 0 & 1 & 1 & 0 & 0 \\
    0 & 1 & 0 & 0 & 1 & 1\\
    0 & 1 & 0 & 0 & 1 & 1 }\quad.
\end{equation*}
It was shown in \cite{Candelas:2008wb} that in the first form, $X^{19,19}$ admits a
free action of the order-eight quaternion group (denoted in \cite{Candelas:2008wb} by
$\IH$, but more conventionally by $Q_8$), with elements $\{\pm 1, \pm \ii, \pm j, \pm k\}$,
induced by a linear action of this group on the ambient space.

In the second form, $X^{19,19}$ admits free actions by two groups of order twelve.  One
is the cyclic group $\IZ_{12}$, and the other is the dicyclic group
$\Dic_3 \cong \IZ_3{\rtimes}\IZ_4$ (introduced in \sref{sec:(8,44)}) \cite{Davies:DPhil}.  These
were in fact discovered via conifold transitions from the corresponding quotients of
$X^{8,44}$, an idea reviewed in \sref{sec:transitions}.
\begin{table}[ht]
\begin{center}
    \begin{tabular}{| c  c  c |}
        \hline
        \str
        $~h^{1,1} = h^{2,1}~$ & ~Fundamental Group~ & ~Reference~ \\
        \hline
        \hline
        11  &  $\IZ_2$  &  \cite{Candelas:2008wb,Bouchard:2007mf} \\
        \hline
        7  &  $\IZ_3~,~\IZ_2{\times}\IZ_2$  &  \cite{Candelas:2008wb,Bouchard:2007mf} \\
        \hline
        5  &  $\IZ_4$  &  \cite{Bouchard:2007mf} \\
        \hline
        3  &  $\IZ_5~,~\IZ_6~,~\IZ_4{\times}\IZ_2~,~Q_8~,~\IZ_3{\times}\IZ_3$  &  \cite{Candelas:2008wb,Bouchard:2007mf} \\
        \hline
        2  &  $\IZ_{12}~,~\Dic_3$  &  \cite{Braun:2009qy,Davies:DPhil} \\
        \hline
    \end{tabular}\\
    \parbox{.75\textwidth}{\caption{\label{tab:1919hodgenos}
    \small  The Hodge numbers for known quotients of $X^{19,19}$.
    }}
\end{center}
\vskip-10pt
\end{table}

In summary, $X^{19,19}$ is rather exceptional in that it admits free actions by the groups
$\IZ_{12}, \Dic_3, \IZ_3{\times}\IZ_3, \IZ_4{\times}\IZ_2, Q_8, \IZ_6$ and $\IZ_5$.  The
Hodge numbers for the quotients by all these groups and their subgroups are collected
in \tref{tab:1919hodgenos}.

\subsubsection{Manifolds with Hodge numbers (1,1)} \label{sec:(1,1)}

For a long time, the smallest known Hodge numbers of a Calabi-Yau threefold
satisfied $h^{1,1} + h^{2,1} = 4$.  This record has now been overtaken by Braun's
examples of manifolds with $\hodgenos = (1,1)$ \cite{Braun:2011hd} (as well
as Freitag and Salvati Manni's manifold with $\hodgenos = (2,0)$
\cite{Freitag:2011st}).

The covering space of Braun's $(1,1)$ manifolds is a self-mirror manifold
$X^{20,20}$.  This is realised as an anti-canonical hypersurface in the toric fourfold
determined by the face fan over the 24-cell, which is a self-dual regular
four-dimensional polytope.

There are three different groups of order 24 which act freely on particular smooth
one-parameter sub-families of $X^{20,20}$; these are
$\IZ_3{\rtimes}\IZ_8, \IZ_3{\times}Q_8$ and $SL(2,3)$.  The first two are
self-explanatory, while the third is the group of two-by-two matrices of determinant one
over the field with three elements.  All the groups act via linear transformations on the
lattice in which the polytope lives, and act transitively on its vertices.
Full details can be found in \cite{Braun:2011hd}.

\subsubsection{Complete intersections of four quadrics in
\texorpdfstring{$\IP^7$}{P7}} \label{sec:fourquadrics}

A particularly fertile starting point for finding new Calabi-Yau manifolds has been the
complete intersection of four quadrics in $\IP^7$.  A smooth member of this family is
a Calabi-Yau manifold with Hodge numbers $\hodgenos = (1,65)$.  Hua classified
free group actions on smooth sub-families in \cite{Hua}, finding groups of order 2, 4,
8, 16 and 32.  The quotients all have $h^{1,1} = 1$, and $h^{2,1} =$ 33, 17, 9, 5 and
3 respectively.

Certain nodal families allow free actions of groups of order 64, and furthermore have
equivariant small resolutions \cite{GrossPopescuI,Hua}.  The resolutions have Hodge
numbers $\hodgenos = (2,2)$, and inherit the free group actions.  Remarkably, in
this case all the quotients have the same Hodge numbers as the covering space.
The quotient by $\IZ_8{\times}\IZ_8$ was investigated as a background for heterotic
string theory in \cite{Bak:2008ey}, but unfortunately no realistic models were found.

Freitag and Salvati Manni have also constructed a large number of new manifolds by
starting with a particular complete intersection $X^\sharp$ of four quadrics which has
96 nodes and a very large symmetry group \cite{Freitag:2010st,Freitag:2011st}.  They
show that the quotients by many subgroups admit crepant projective resolutions,
thereby giving rise to a large number of new Calabi-Yau manifolds.  Some of the
subgroups of order 2, 4, 8 and 16 act freely on a small resolution of $X^\sharp$, and
the corresponding quotient manifolds are connected to some of Hua's examples by
conifold transitions.  The manifolds from \cite{Freitag:2011st} with small Hodge
numbers are listed in \aref{app:new}, including one with $\hodgenos = (2,0)$, which
is therefore equal with Braun's manifolds for smallest known Hodge numbers.  Note
that the theoretical minimum is $\hodgenos = (1,0)$.

\section{New manifolds from topological transitions} \label{sec:transitions}

One fascinating feature of Calabi-Yau threefolds is the inter-connectedness of moduli
spaces of topologically distinct manifolds.  Generally speaking, there are two ways to
pass from one smooth Calabi-Yau to another.  We may deform the complex structure
until a singularity develops, and then `resolve' this singularity using the techniques of
algebraic geometry, which involves replacing the singular set with new embedded
holomorphic curves or surfaces.  Alternatively, we may allow certain embedded curves
or surfaces to collapse to zero size, and then `smooth' the resulting singular space by
varying its complex structure.  Obviously these two processes are inverses of each
other.

Our main interest here is in constructing new smooth Calabi-Yau threefolds via
such topological transitions, but first I will indulge in a few comments about the
connectedness of the space of all Calabi-Yau threefolds.

The suggestion that all Calabi-Yau threefolds might be connected by topological
transitions goes back to \cite{ReidFantasy}.  Early work showed
that this was true for nearly all examples known at the time \cite{Green:1988bp,
Green:1988wa}.  These papers considered conifold transitions, in which the intermediate
variety has only nodal singularities; the smoothing process replaces these singular points
by three-spheres, while the `small' resolution replaces them by two-spheres
(holomorphically embedded).  Such singularities were shown to be at finite distance in
moduli space \cite{Candelas:1988di,Candelas:1989ug}, and conifold transitions were
later shown to be smooth processes in type II string theory \cite{Strominger:1995cz,
Greene:1995hu}.

If we wish to connect all Calabi-Yau threefolds, conifold transitions are not sufficient,
because they cannot change the fundamental group.  To see this we note that,
topologically, a conifold transition consists of removing neighbourhoods of some
number of copies of $S^3$, each with boundary $S^3{\times}S^2$, and replacing
them with similar neighbourhoods of $S^2$.  Since all these spaces are
simply-connected, a simple application~of van Kampen's theorem (see e.g.
\cite{Hatcher}) shows that the fundamental group does not change.

There do exist relatively mild topological transitions which can change the
fundamental group; these are known as \emph{hyperconifold transitions}, and
were described by the author in \cite{Davies:2009ub,Davies:2011is}.  Here the
singularities of the intermediate variety are finite quotients of a node, and
their resolutions are no longer `small'.  It is an interesting question whether all
Calabi-Yau threefolds can be connected by conifold and hyperconifold transitions.

In the following sections, we will consider these two types of transition separately,
mostly through examples.  The examples in \sref{sec:hyperconifolds} actually yield
previously unknown manifolds, with Hodge numbers $\hodgenos = (2,5)$ and
$(2,3)$ and fundamental groups $\IZ_5$ and $S_3$ respectively.

\subsection{Conifold transitions} \label{sec:conifolds}

In \cite{Candelas:2008wb,Candelas:2010ve}, free group actions were followed through
conifold transitions, leading to webs of conifold transitions between smooth quotients
with the same fundamental group (conifold transitions were also used in
\cite{Batyrev:2008rp,Filippini:2011rf} to construct new simply-connected manifolds).
Here we will just consider a simple example (taken from \cite{Candelas:2008wb}) which
exemplifies the idea.

Consider the well-known family of quintic hypersurfaces in $\IP^4$, with Hodge numbers
$\hodgenos = (1,101)$ and hence Euler number $\chi = -200$.  If we take
homogeneous coordinates $z_0,\ldots,z_4$ on $\IP^4$, then an action of $\IZ_5$
can be defined by the generator
\begin{equation*}
    g_5 : z_i \to z_{i+1} ~.
\end{equation*}
Then there is a smooth family of invariant quintics, given by
\begin{equation} \label{eq:symquintic}
    f = \sum_{ijklm} \, A_{j-i,k-i,l-i,m-i}\, z_i z_j z_k z_l z_m = 0~.
\end{equation}
For generic coefficients, $\IZ_5$ acts freely, so we get a family of smooth quotients with
Hodge numbers $\hodgenos = (1,21)$.  

Now let us consider a non-generic choice for the coefficients in \eqref{eq:symquintic},
such that $f$ is the determinant of some $5\times5$ matrix $M$ which is linear in the
homogeneous coordinates.  If we take the entries
of $M$ to be
\begin{equation*}
    M_{ik} = \sum_j \, a_{j-i,k-i}\, z_j ~,
\end{equation*}
then the induced $\IZ_5$ action is $M_{ik} \to M_{i+1, k+1}$, so the determinant
does indeed correspond to an invariant quintic.  The action of $\IZ_5$ is still generically
fixed-point-free on the family given by $\det M = 0$, but the hypersurfaces are no longer
smooth.  Indeed, using a computer algebra package, it can be checked that the rank of
$M$ drops to three at exactly fifty points on a typical such hypersurface, and that these
points are nodes.  Furthermore, they fall into ten orbits of five nodes under the $\IZ_5$
action.

We now ask whether these nodes can be resolved in a group-invariant way; if so, the
group will still act freely on the resolved manifolds, and we will have constructed a
conifold transition between the quotient manifolds.  In fact this is easy to do.  Introduce
a second $\IP^4$, with homogeneous coordinates $w_0,\ldots,w_4$, and consider the
equations
\begin{equation} \label{eq:splitquintic}
    f_i := \sum_k \, M_{ik}\, w_k = \sum_{j,k} \, a_{j-i, k-i}\, z_j\, w_k = 0 ~.
\end{equation}
These are five bilinears in $\IP^4{\times}\IP^4$, and it can be checked that they generically
define a smooth Calabi-Yau threefold $X^{2,52}$.  Since we cannot have
$w_i = 0 ~\forall~ i$, there are only simultaneous solutions to these equations when
$\det M = 0$, so this gives a projection from $X^{2,52}$ to nodal members of $X^{1,101}$.
At most points, this is one-to-one, but at the fifty points where the rank of $M$ drops to
three, we get a whole copy of $\IP^1 \subset \IP^4$ projecting to a (nodal) point of
$X^{1,101}$.  In this way we see that we have constructed a conifold transition
$X^{1,101} \rightsquigarrow X^{2,52}$.

To see that the free $\IZ_5$ action is preserved by the conifold transition above, it
suffices to note that if we extend the action by defining $g_5 : w_i \to w_{i+1}$, then
this induces $f_i \to f_{i+1}$, implying that the manifolds defined by \eqref{eq:splitquintic}
are $\IZ_5$-invariant.  The absence of fixed points follows from the absence of fixed
points on the nodal members of $X^{1,101}$, although this can also be checked directly.

Since the conifold transition from $X^{1,101}$ to $X^{2,52}$ can be made
$\IZ_5$-equivariant, it descends to a conifold transition between their quotients,
$X^{1,21} \rightsquigarrow X^{2,12}$, where the intermediate variety has ten nodes.

\subsection{Hyperconifold transitions} \label{sec:hyperconifolds}

The conifold transition in the last section illustrates two completely general features of
such transitions:  the fundamental group does not change, for reasons explained
previously, and the intermediate singular variety has multiple nodes
\cite{Candelas:1989js}.  We now turn our attention to a class of transitions for which
neither of these statements hold --- the so-called hyperconifold transitions introduced
in \cite{Davies:2009ub}.  Here the intermediate space typically has only one singular
point, which is a quotient of a node by some finite cyclic group
$\IZ_N$.\footnote{Quotients by non-Abelian groups can also occur, but these do not
admit a toric description, and their resolutions have not been studied.}  These arise
naturally when a generically-free group action is allowed to develop a fixed point.  A
$\IZ_N$-hyperconifold transition changes the Hodge numbers according to the general
formula
\vspace{-5pt}
\begin{equation} \label{eq:hchodge}
    \d\hodgenos = (N-1, -1) ~.
\end{equation}
\vspace{-25pt}

The resolution of a hyperconifold singularity replaces the singular point with a
simply-connected space, and in this way we see that the transitions can change the
fundamental group.  It is worth pausing here to consider this in more detail than has
been done in previous papers.

Suppose we have a smooth quotient $X = \widetilde X/G$, and deform the complex
structure until some order-$N$ element $g_N$, which generates a subgroup
$\langle g_N \rangle \cong \IZ_N\! < G$, develops a fixed point $p \in \widetilde X$.
Then, as described in \cite{Davies:2009ub}, this point will be singular, and generically
a node.  In some cases, the group structure implies that other elements will
simultaneously develop fixed points, which we can see by taking a group element
$g' \in G\setminus\langle g_N\rangle$ and performing an elementary calculation,
\begin{equation*}
    g'\,g_N\,g'^{-1} \cdot (g'\cdot p) = g'\cdot(g_N\cdot p) = g'\cdot p ~.
\end{equation*}
So the point $g' \cdot p \in \widetilde X$ is fixed by $g'\,g_N\,g'^{-1}$.  We see that every
subgroup conjugate to $\langle g_N \rangle$ also develops a fixed point.  All such
points are identified by $G$, so the singular quotient $X^\sharp$ has only one
$\IZ_N$-hyperconifold singularity.

What is the fundamental group of the resolution $\widehat X^\sharp$?  To calculate this,
excise a small ball around each fixed point of $\widetilde X^\sharp$, to obtain a smooth
space $\widetilde X'$ on which the whole group $G$ acts freely.  We can then quotient
by $G$ to obtain $X'$, with fundamental group $G$.  Finally, we glue in a
neighbourhood $\S$ of the exceptional set of the resolution of the hyperconifold.  $\S$
is simply-connected.  We now have $\widehat X^\sharp = X' \cup \S$, and can use van
Kampen's theorem to calculate $\pi_1(\widehat X^\sharp)$.  Note that the intersection
of the two subspaces $X'$ and $\S$ is homotopy-equivalent to $S^3{\times}S^2/\IZ_N$,
since the stabiliser of each point on the covering space was isomorphic to $\IZ_N$.  So
we have the data
\begin{equation*}
    \widehat X^\sharp = X' \cup \S ~,~~ X' \cap \S \stackrel{\mathrm{hom.}}{\simeq} S^3{\times}S^2/\IZ_N 
        ~,~~ \pi_1(\S) \cong \id ~,~~ \pi_1(X') \cong G ~,
\end{equation*}
which by van Kampen's theorem immediately implies that
$\pi_1(\widehat X^\sharp) \cong G/\langle g_N \rangle^G$, where $\langle g_N \rangle^G$
is the smallest normal subgroup of $G$ which contains $\langle g_N \rangle$, usually
called the normal closure.

Trivial examples arise when $G = \IZ_N{\times}H$ or $\IZ_N{\rtimes}H$, and
the generator of $\IZ_N$ develops a fixed point.  In this case, the corresponding
hyperconifold transition changes the fundamental group from $G$ to $H$.

\subsubsection{Example 1:
\texorpdfstring{$X^{1,6} \rightsquigarrow X^{2,5}$}{X(1,6) to X(2,5)}} \label{sec:ex1}

We will first consider an example related to that in \sref{sec:conifolds}.  If we demand that
the matrix appearing in \eqref{eq:splitquintic} is symmetric, $a_{jk} = a_{kj}$, then the
resulting family of threefolds are invariant under a further order-two symmetry, generated
by $g_2 : z_i \longleftrightarrow w_i$.  As shown in \cite{Candelas:2008wb}, this family
is still generically smooth, and the entire group $\IZ_5{\times}\IZ_2~\cong~\IZ_{10}$
acts freely, so we get a smooth quotient family $X^{1,6}$.

Suppose now that we ask for $g_2$ to develop a fixed point.  In the ambient space,
it fixes an entire copy of $\IP^4$, given by $w_i = z_i ~\forall~i$.  Choose a single
point on this locus (as long as it is not also a fixed point of $g_5$), say
$w_i = z_i = \d_{i0}$.  The evaluation of the defining polynomials at this point is
$f_i = c_{-i,-i}$, so it lies on the hypersurface if $c_{i,i} = 0 ~\forall~i$.  One can check
that for arbitrary choices of the other coefficients, this point is a node on the covering
space, and there are no other singularities.  This therefore corresponds to a sub-family
of $X^{1,6}$ with an isolated $\IZ_2$-hyperconifold singularity.  Such a singularity has a
crepant projective resolution, as described in \cite{Davies:2009ub}, obtained by a simple
blow-up of the singular point.  This introduces an irreducible exceptional divisor, thus
increasing $h^{1,1}$ by one, and since we imposed a single
constraint\footnote{Na\"ively, it seems that we have imposed five constraints,
$c_{i,i} = 0$.  However, we had the freedom to choose a generic point on the fixed
$\IP^4$, corresponding to a four-parameter choice of possible conditions, so the number
of complex structure parameters is actually only reduced by one.} on the complex
structure of $X^{1,6}$, the resolved space has Hodge numbers $\hodgenos = (2,5)$,
and its fundamental group is $\IZ_5$.\footnote{Note that this is in fact a new
`three-generation' manifold, with $\chi = -6$.  Unfortunately, $\IZ_5$-valued Wilson
lines cannot perform the symmetry breaking required for a realistic model.}

So we have constructed a $\IZ_2$-hyperconifold transition
$X^{1,6} \stackrel{\IZ_2}{\rightsquigarrow} X^{2,5}$, where the fundamental group of
the first space is $\IZ_{10}$, and that of the second space is $\IZ_5$.

\subsubsection{Example 2:
\texorpdfstring{$X^{1,4} \rightsquigarrow X^{2,3}$}{X(1,4) to X(2,3)}} \label{sec:ex2}

For a second example, which will also yield an interesting new manifold, consider the
$\Dic_3$ quotient of $X^{8,44}$, described in \sref{sec:(8,44)}.  As shown in
\cite{Braun:2009qy}, there is a codimension-one locus in moduli space where the
unique order-two element of the group develops a fixed point.  It is easy to check that
on the covering space, this is the only singular point, and is a node.  As such, the
quotient space $X^{1,4}$ develops a $\IZ_2$-hyperconifold singularity.  Blowing up
this point yields a new Calabi-Yau manifold, with Hodge numbers $\hodgenos = (2,3)$,
as per the general formula \eqref{eq:hchodge}.

The $\IZ_2$ subgroup of $\Dic_3$ is actually the centre, so it is trivially normal, and the
fundamental group of the new manifold $X^{2,3}$ is $\Dic_3/\IZ_2$, which is isomorphic
to $S_3$, the symmetric group on three letters.  To see this, recall that $\Dic_3$ is
generated by two elements, $g_3$ and $g_4$, of orders three and four respectively,
subject to the relation $g_4 g_3 g_4^{-1} = g_3^2$.  So the $\IZ_2$ subgroup is
generated by $g_4^2$, meaning that in $\Dic_3/\IZ_2$, $g_4^2 \sim e$.  To reflect
this, we rename $g_4$ to $g_2$, and obtain
\begin{equation*}
    \Dic_3/\IZ_2 ~\cong~ \langle\, g_2, g_3 ~\big\vert~ g_2^2 = g_3^3 = e ~,~ g_2 g_3 g_2 = g_3^2 \,\rangle~,
\end{equation*}
which is the standard presentation of $S_3$.

So in summary, we have constructed a $\IZ_2$-hyperconifold transition
$X^{1,4} \stackrel{\IZ_2}{\rightsquigarrow} X^{2,3}$, where the fundamental group of the
first space is $\Dic_3$, and that of the second space is $S_3$.  This is the first known
Calabi-Yau threefold with fundamental group $S_3$ \cite{Braun:2010vc}.

\subsection*{Acknowledgements}

I would like to thank Eberhard Freitag for useful correspondence, and Philip Candelas
for helpful comments on a draft of this paper.  This work was supported by the
Engineering and Physical Sciences Research Council [grant number EP/H02672X/1].

\appendix

\section{The new-look zoo} \label{app:new}

\begin{figure}[ht]
\begin{center}
    \includegraphics[width=.75\textwidth]{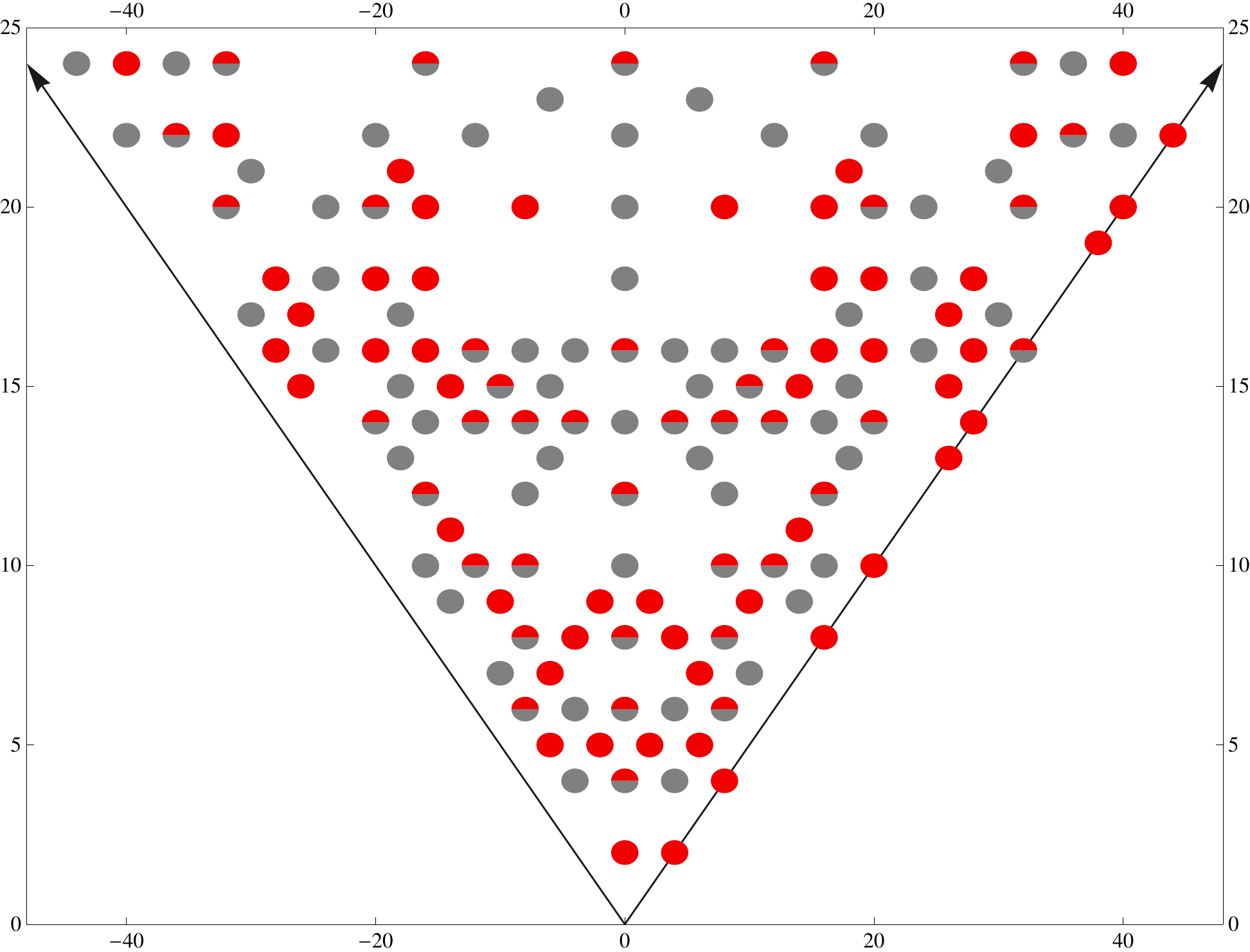}\\[2.5ex]
    \parbox{.75\textwidth}{\caption{\label{fig:tip}
    \small
    The tip of the distribution of Calabi-Yau threefolds.  Grey dots denote manifolds
    included in \cite{Candelas:2008wb}, while red dots denote newer examples.
    Split dots indicate multiple occupation of a site.  Note that some red and grey dots
    are also multiply-occupied.
    }}\\
    \place{0.1}{3.3}{\footnotesize$h^{1,1}+h^{2,1}$}
    \place{2.6}{1.05}{\footnotesize$\chi = 2(h^{1,1}-h^{2,1})$}
\end{center}
\end{figure}

The techniques reviewed in Sections \ref{sec:quotients} and \ref{sec:transitions}, along
with a few exceptional constructions, have led in recent years to the construction of a
relatively large number of new Calabi-Yau threefolds with small Hodge numbers and/or
non-trivial fundamental group.  A table appeared in \cite{Candelas:2008wb} of all
manifolds known at the time with $h^{1,1} + h^{2,1} \leq 24$.  Instead of repeating that
list here, only new manifolds discovered since the appearance of \cite{Candelas:2008wb}
are listed in \tref{tab:multtip} and \tref{tab:simptip}.  Since they are of most relevance for
string theory, those with non-trivial fundamental group are listed separately in
\tref{tab:multtip}, while \tref{tab:simptip} contains new simply-connected manifolds and
those with fundamental group yet to be calculated.  Figure \ref{fig:tip} displays the tip of
the distribution of manifolds catalogued by their Hodge numbers, showing which values
of $(h^{1,1},h^{2,1})$ satisfying $h^{1,1} + h^{2,1} \leq 24$ are realised by known
examples (and their mirrors, which are assumed to exist).
%

\newlength{\myht}
\newlength{\mydp}
\newlength{\mywd}
\newsavebox{\mybox}

\newcommand{\entry}[2]{\settowidth{\mywd}{\footnotesize${}\quotient{#2}$}%
\hspace{\mywd}%
\sbox{\mybox}{\footnotesize$#1\quotient{#2}$}%
\settoheight{\myht}{\usebox{\mybox}}\addtolength{\myht}{6pt}%
\settodepth{\mydp}{\usebox{\mybox}}\addtolength{\mydp}{5pt}%
\vrule height\myht width0pt depth\mydp\usebox{\mybox}}

\newcommand{\simpentry}[1]{%
\sbox{\mybox}{$#1$}%
\settoheight{\myht}{\usebox{\mybox}}\addtolength{\myht}{6pt}%
\settodepth{\mydp}{\usebox{\mybox}}\addtolength{\mydp}{5pt}%
\vrule height\myht width0pt depth\mydp\usebox{\mybox}}
\newpage
\begin{center}

\setlength{\doublerulesep}{3pt}

\begin{longtable}{| c | c | c | c | c |}

\caption{\label{tab:multtip} Manifolds with small Hodge numbers and $\pi_1 \neq \id$}\\
\multicolumn{5}{c}{\parbox{6.3in}{\small This table complements the one in \cite{Candelas:2008wb}, and
briefly describes the manifolds which have $y~=~h^{1,1}{+}h^{2,1}\leq 24$ and non-trivial fundamental group
discovered since that paper appeared in 2008.  There should still be a number of other manifolds in
this region, including quotients from \cite{Braun:2010vc} whose Hodge numbers have not yet been
calculated, and manifolds obtained from known quotients by hyperconifold transitions
\cite{Davies:2011is}, of which only a few have so far been written down explicitly.
In the `Manifold' column, $X^{20,20}$ denotes the Calabi-Yau toric hypersurface associated to the
24-cell, discussed in \cite{Braun:2011hd} and \sref{sec:(1,1)}, while $X^{19,19}$ refers to the manifold
discussed in \sref{sec:(19,19)}, and $X^{8,44}$ to that in \sref{sec:(8,44)}. $\text{dP}_n$ is the del Pezzo
surface of degree $n$. Multiple quotient groups
indicate different quotients with the same Hodge numbers.  $X^\sharp$ denotes a singular member of
a generically-smooth family, while $\widehat{X}$ denotes a resolution of a singular variety $X$.  The
column labelled by $\pi_1$ gives the fundamental group.  For each manifold listed here there should
also be a mirror, which is not listed.}}
\vspace{10pt}\\
\hline 
\str $(\chi,\, y)$ & $\!(h^{1,1},h^{2,1})\!\!$ & Manifold & $\pi_1$ & Reference \\ 
\hline\hline
\endfirsthead

\multicolumn{5}{c}{\tablename\ \thetable{} \emph{-- Continued from previous page}} \\[1ex]
\hline 
\str $(\chi,\, y)$ & $(h^{11},h^{21})$ & Manifold & $\pi_1$ & Reference \\
\hline\hline 
\endhead

\hline\hline
\multicolumn{5}{|r|}{\str\emph{Continued on the following page}}\\
\hline
\endfoot

\hline\hline
\multicolumn{5}{|c|}{\str}\\
\hline
\endlastfoot
\str
(0,24) & (12,12) & $X^{20,20}/\IZ_2$ & $\IZ_2$ & \cite{Braun:2011hd}\\
\hline
\str
(-16,18) & (5,13) & (Hypersurface in $\IP^1{\times}\IP^1{\times}\text{dP}_4$)/$\IZ_2{\times}\IZ_2$ & $\IZ_2{\times}\IZ_2$ & \cite{Bini:2011dp}\\
\hline
(-20,16) & (3,13) &
\entry{\cicy{\IP^1 \\ \IP^1\\  \IP^1\\ \IP^2\\ \IP^2\\ \IP^2}
{ 1~ 0~ 0~ 1 ~ 0 ~ 0\\
 0~ 1~ 0~  1 ~ 0 ~ 0\\
 0~ 0~ 1~ 1 ~ 0 ~ 0\\
 1~ 0~ 0~  0 ~ 1 ~ 1\\
 0~ 1~ 0~  0 ~ 1 ~ 1\\
 0~ 0~ 1~ 0 ~ 1 ~ 1}}{\IZ_3}  & $\IZ_3$ & \cite{Braun:2010vc,Candelas:2010ve} \\
\hline
(-12,16) & (5,11) &
\entry{\cicy{\IP^1 \\ \IP^1\\ \IP^1\\ \IP^3\\ \IP^2}
{ 1~ 1~ 0 ~ 0 ~ 0\\
 1~ 1~  0 ~ 0 ~ 0\\
 1~ 1~ 0 ~ 0 ~ 0\\
 0 ~1~  1 ~ 1 ~ 1\\
 0~ 0~  1 ~ 1 ~ 1}}{\IZ_3}  & $\IZ_3$ & \cite{Braun:2010vc,Candelas:2010ve} \\
\hline
\str
(0,16) & (8,8) & $X^{20,20}/\IZ_3$ & $\IZ_3$ & \cite{Braun:2011hd}\\
\hline
\str
(0,16) & (8,8) & (Toric hypersurface $Y^{20,20})/\IZ_3$ & $\IZ_3$ & \cite{Davies:2011is}\\
\hline
\Str
(32,16) & (16,0) & $\widehat{\left(\IP^7[2~2~2~2]^\sharp/\IZ_2\right)^{\phantom\sharp}}$ & $\IZ_2$ & \cite{Freitag:2011st} \\
\hline
(-14,15) & (4,11) &
\entry{\cicy{\IP^1 \\ \IP^1\\ \IP^1\\ \IP^2\\ \IP^2\\ \IP^2\\ \IP^2}
{ 1~ 0~ 0~ 0~0~ 0~ 1 ~ 0\\
 0~ 1~ 0~ 0~0~ 0~ 1 ~ 0\\
 0~ 0~ 1~ 0~0~ 0~ 1 ~ 0\\
1~ 0~ 0~ 1~ 0~ 0~ 0 ~ 1\\
0~ 1~ 0~ 0~ 1~ 0~ 0 ~ 1\\
 0~ 0~ 1~0~ 0~ 1~ 0 ~ 1\\
0~ 0~ 0~1~1~ 1~ 0 ~ 0\\}}{\IZ_3}  & $\IZ_3$ & \cite{Braun:2010vc,Candelas:2010ve} \\
\hline
(-10,15) & (5,10) &
\entry{\cicy{\IP^1 \\ \IP^1\\ \IP^1\\ \IP^1\\ \IP^1\\ \IP^1\\ \IP^2\\ \IP^2\\  \IP^2}
{ 1~ 0~ 0~ 0~ 0 ~ 0~ 1~0~ 0  \\
 0~ 1~ 0~ 0 ~ 0 ~ 0 ~ 1~ 0~ 0 \\
 0~ 0~ 1~ 0~ 0~0~ 1 ~ 0~ 0 \\
 0~ 0~ 0~ 1~ 0~0~ 0 ~ 1~ 0\\
 0~ 0~ 0~ 0~ 1~ 0~ 0 ~ 1~ 0 \\
 0~ 0~ 0~ 0~ 0~ 1 ~  0 ~ 1~ 0 \\
 1~ 0~ 0~ 1~0 ~0~ 0~0 ~ 1\\
 0~ 1~ 0~ 0~ 1~0~ 0~ 0 ~ 1\\
 0~ 0~ 1~ 0~ 0~1~ 0~0 ~ 1\\}}{\IZ_3}  & $\IZ_3$ & \cite{Braun:2010vc,Candelas:2010ve} \\
\hline
\str
(-12,14) & (4,10) & (Toric hypersurface $X^{8,26}$)/$\IZ_3$ & $\IZ_3$ & \cite{Davies:2011is} \\
\hline
(-8,14) & (5,9) &
\entry{\cicy{\IP^1 \\ \IP^1\\ \IP^1\\ \IP^2\\ \IP^2\\ \IP^2\\ \IP^2\\ \IP^2}
{ 1~ 0~ 0~ 0~ 0~ 0~0~ 0~ 0~ 1 \\
 0~ 1~ 0~ 0~ 0~0~0~ 0~ 0~ 1 \\
 0~ 0~ 1~ 0~ 0~0~0~0~ 0~ 1 \\
1 ~ 0~ 0~ 1~ 0~0~1~ 0~ 0~ 0 \\
 0~ 1 ~ 0~ 0~ 1~ 0~ 0~ 1~ 0~ 0 \\
 0~ 0~ 1~ 0~ 0~ 1 ~0~0~ 1~ 0 \\
0~ 0~ 0 ~ 1~ 1 ~1 ~ 0~ 0~ 0~0 \\
0~ 0~ 0~ 0~ 0~0~1~ 1~ 1~ 0 \\}}{\IZ_3} & $\IZ_3$ & \cite{Braun:2010vc,Candelas:2010ve} \\
\hline
(-4,14) & (6,8) &
\entry{\cicy{\IP^1 \\ \IP^1\\ \IP^1\\ \IP^1\\ \IP^1\\ \IP^1\\ \IP^2\\ \IP^2\\ \IP^2\\ \IP^2}
{1~ 0~ 0~ 0~ 0 ~ 0~0~0~ 0~ 1~ 0 \\
 0~ 1~ 0~ 0 ~ 0 ~ 0 ~0~0~ 0~ 1~ 0 \\
 0~ 0~ 1~ 0~ 0~0~0~0~ 0~ 1~ 0 \\
 0~ 0~ 0~ 1~ 0~0~0~0~ 0~ 0~ 1 \\
 0~ 0~ 0~ 0~ 1~ 0~ 0~0 ~ 0~ 0~ 1 \\
 0~ 0~ 0~ 0~ 0~ 1 ~0~0~ 0~ 0~ 1 \\
1~ 0~ 0~ 1~0 ~0~1~ 0~ 0~ 0~0 \\
0~ 1~ 0~ 0~ 1~0~0~ 1~ 0~ 0~ 0 \\
0~ 0~ 1~ 0~ 0~1~0~ 0~ 1~ 0~0 \\
0~ 0~ 0~ 0~ 0~0~1~1~ 1~ 0~ 0}}{\IZ_3}  & $\IZ_3$ & \cite{Braun:2010vc,Candelas:2010ve} \\
\hline
\str
(0,12) & (6,6) & $X^{20,20}/\IZ_4$ & $\IZ_4$ & \cite{Braun:2011hd}\\
\hline
\str
(-10,9) & (2,7) & (Hypersurface in $\text{dP}_5{\times}\text{dP}_5$)/$\IZ_5$ & $\IZ_5$ & \cite{Bini:2011dp}\\
\hline
\str
(2,9) & (5,4) & (Toric hypersurface $X^{21,16})/\IZ_5$ & $\IZ_5$ & \cite{Davies:2011is}\\
\hline
\str
(-4,8) & (3,5) & (Hypersurface in $\text{dP}_4{\times}\text{dP}_4$)/$\IZ_4{\times}\IZ_2$ & $\IZ_4{\times}\IZ_2$ & \cite{Bini:2011dp}\\
\hline
\str
(0,8) & (4,4) & $X^{20,20}/\IZ_6$ & $\IZ_6$ & \cite{Braun:2011hd}\\
\hline
\Str
(16,8) & (8,0) & $\widehat{\left(\IP^7[2~2~2~2]^\sharp/\{\IZ_2{\times}\IZ_2,\IZ_4\}\right)^{\phantom\sharp}} $ &
    $\IZ_2{\times}\IZ_2,\,\IZ_4$ & \cite{Freitag:2011st} \\
\hline
\Str
(-6,7) & (2,5) & $\widehat{\left({X^{2,52}}/\IZ_{10}\right)^{\displaystyle\sharp}}$ & $\IZ_5$ & \!\sref{sec:ex1}\!\! \\
\hline
\str
(-8,6) & (1,5) & \big(Hypersurface in $(\IP^1)^4$\big)/$\IZ_8{\times}\IZ_2$ & $\IZ_8{\times}\IZ_2$ & \cite{Bini:2011dp}\\
\hline
\str
(0,6) & (3,3) & $X^{20,20}/\{\IZ_8, Q_8\}$ & $\IZ_8, Q_8$ & \cite{Braun:2011hd}\\
\hline
\str
(-6,5) & (1,4) & $X^{8,44}/\{\Dic_3, \IZ_{12}\}$ & $\Dic_3, \IZ_{12}$ & \cite{Braun:2009qy}\\
\hline
\Str
(-2,5) & (2,3) & $\widehat{\left({X^{8,44}}/\Dic_3\right)^{\displaystyle\sharp}}$ & $S_3$ & \!\sref{sec:ex2}\!\! \\
\hline
\str
(0,4) & (2,2) & $X^{19,19}/\{\Dic_3,\IZ_{12}\}$ & $\Dic_3, \IZ_{12}$ & \cite{Braun:2009qy,Davies:DPhil}\\
\hline
\str
(0,4) & (2,2) & $X^{20,20}/\IZ_{12}$ & $\IZ_{12}$ & \cite{Braun:2011hd}\\
\hline
\Str
(8,4) & (4,0) & $\widehat{\left(\IP^7[2~2~2~2]^\sharp/G\right)^{\phantom\sharp}} ~,~|G|=8$ & $G$ & \cite{Freitag:2011st} \\
\hline
\str
(0,2) & (1,1) & $X^{20,20}/\{SL(2,3), \IZ_3\rtimes\IZ_8, \IZ_3\times Q_8\}$ &
    \hspace{-5pt}$\begin{array}{ll} SL(2,3), & \!\!\!\!\IZ_3\rtimes\IZ_8, \\ \IZ_3\times Q_8 & \end{array}\hspace{-5pt}$ & \cite{Braun:2011hd}\\
\hline
\Str
(4,2) & (2,0) & $\widehat{\left(\IP^7[2~2~2~2]^\sharp/G\right)^{\phantom\sharp}} ~,~|G|=16$ & $G$ & \cite{Freitag:2011st} \\

\end{longtable}
\end{center}

\newpage

\begin{center}
\begin{longtable}{| c | c | c | c |}

\caption{\label{tab:simptip} Other manifolds with small Hodge numbers}\\
\multicolumn{4}{c}{\parbox{6.3in}{\small This table is the same as that above, except all the
manifolds listed either have trivial fundamental group, or a fundamental group which has
not been calculated (which is the case for several examples from \cite{Freitag:2011st}).
The notation is the same as above, and the manifolds with no description are all
desingularisations of quotients by various groups of a singular complete intersection
of four quadrics in $\IP^7$ \cite{Freitag:2011st}.}}
\vspace{10pt}\\
\hline 
\str $(\chi,\, y)$ & $(h^{1,1},h^{2,1})$ & Manifold & Reference \\ 
\hline\hline
\endfirsthead

\multicolumn{4}{c}{\tablename\ \thetable{} \emph{-- Continued from previous page}} \\[1ex]
\hline 
\str $(\chi,\, y)$ & $(h^{11},h^{21})$ & Manifold & Reference \\
\hline\hline 
\endhead

\hline\hline
\multicolumn{4}{|r|}{\str\emph{Continued on the following page}}\\
\hline
\endfoot

\hline\hline
\multicolumn{4}{|c|}{\str}\\
\hline
\endlastfoot
\str
(16,24) & (16,8) & --- & \cite{Freitag:2011st} \\
\hline
\str
(32,24) & (20,4) & --- & \cite{Freitag:2011st} \\
\hline
\str
(40,24) & (22,2) & --- & \cite{Freitag:2011st} \\
\hline
\Str
(-32, 22) & (3,19) & $\widehat{\left(\IP^4[5]/D_5\right)^{\phantom\sharp}}$ & \cite{Stapledon:2010mo} \\
\hline
(36,22) & (20,2) & \simpentry{\parbox{2.2in}{Smoothing of variety obtained
								  by blowing down 18 rational curves on
								  the rigid `$Z$' manifold.}} & \cite{Filippini:2011rf} \\
\hline
\str
(44,22) & (22,0) & --- & \cite{Freitag:2011st} \\
\hline
(18,21) & (15,6) & \simpentry{\parbox{2.2in}{Smoothing of variety obtained
								  by blowing down 27 rational curves on
								  the rigid `$Z$' manifold.}} & \cite{Filippini:2011rf} \\
\hline
\Str
(-20, 20) & (5,15) & $\widehat{\left(\IP^4[5]/A_5\right)^{\phantom\sharp}}$ & \cite{Stapledon:2010mo} \\
\hline
\str
(8,20) & (12,8) & --- & \cite{Freitag:2011st} \\
\hline
\str
(16,20) & (14,6) & --- & \cite{Freitag:2011st} \\
\hline
\str
(32,20) & (18,2) & --- & \cite{Freitag:2011st} \\
\hline
\str
(40,20) & (20,0) & --- & \cite{Freitag:2011st} \\
\hline
\str
(38,19) & (19,0) & --- & \cite{Freitag:2011st} \\
\hline
\str
(20,18) & (14,4) & --- & \cite{Freitag:2011st} \\
\hline
\str
(28,18) & (16,2) & --- & \cite{Freitag:2011st} \\
\hline
\str
(26,17) & (15,2) & --- & \cite{Freitag:2011st} \\
\hline
\str
(16,16) & (12,4) & --- & \cite{Freitag:2011st} \\
\hline
\str
(28,16) & (15,1) & --- & \cite{Freitag:2011st} \\
\hline
\str
(32,16) & (16,0) & --- & \cite{Freitag:2011st} \\
\hline
\str
(26,15) & (14,1) & --- & \cite{Freitag:2011st} \\
\hline
\str
(20,14) & (12,2) & --- & \cite{Freitag:2011st} \\
\hline
\str
(28,14) & (14,0) & --- & \cite{Freitag:2011st} \\
\hline
\str
(26,13) & (13,0) & --- & \cite{Freitag:2011st} \\
\hline
\str
(16,12) & (10,2) & --- & \cite{Freitag:2011st} \\
\hline
\str
(14,11) & (9,2) & --- & \cite{Freitag:2011st} \\
\hline
\str
(8,10) & (7,3) & --- & \cite{Freitag:2011st} \\
\hline
\str
(12,10) & (8,2) & --- & \cite{Freitag:2011st} \\
\hline
\str
(20,10) & (10,0) & --- & \cite{Freitag:2011st} \\
\hline
\str
(8,8) & (6,2) & --- & \cite{Freitag:2011st} \\
\hline
\str
(16,8) & (8,0) & --- & \cite{Freitag:2011st} \\
\hline
\str
(8,4) & (4,0) & --- & \cite{Freitag:2011st} \\
\end{longtable}
\end{center}

\newpage

\bibliographystyle{utphys}
\bibliography{references}

\end{document}